\documentclass[reprint,aps,prper,twocolumn]{revtex4-2} 
\usepackage{graphicx}
\usepackage{hyperref}
\usepackage{footnote}
\usepackage{amsmath,amssymb}
\usepackage{multirow}
\usepackage{tabularx}
\usepackage[table]{xcolor}
\usepackage{booktabs} 
\usepackage{longtable}
\usepackage{cleveref}
\usepackage{makecell}
\usepackage{array}
\usepackage{supertabular}
\usepackage{comment}
\usepackage{enumitem} 
\newcommand{\rowgroup}[1]{\hspace{-1em}#1}

\usepackage{enumitem}
\usepackage{pdfpages}

\makeatletter
\AtBeginDocument{\let\LS@rot\@undefined}
\makeatother

\begin{document}

\title{Seeing quantum effects in experiments}
\date{\today}

\author{Victoria Borish}
\email[]{victoria.borish@colorado.edu}
\author{H. J. Lewandowski}

\affiliation{Department of Physics, University of Colorado, Boulder, Colorado 80309, USA}
\affiliation{JILA, National Institute of Standards and Technology and University of Colorado, Boulder, Colorado 80309, USA}

\begin{abstract}
Quantum mechanics is a field often considered very mathematical, abstract, and unintuitive. One way some instructors are hoping to help familiarize their students with these complex topics is to have the students see quantum effects in experiments in undergraduate instructional labs. Here, we present results from an interview study about what it means to both instructors and students to see quantum effects in experiments. We focus on a popular set of quantum optics experiments, and find that students believe they are observing quantum effects and achieving related learning goals by working with these experiments. Although it is not possible to see the quantum phenomena directly with their eyes, students point out different aspects of the experiments that contribute to them observing quantum effects. This often includes seeing the experimental results, sometimes in conjunction with interacting with or understanding part of the experiment. There is additional variation across student achievement of the various related learning goals, ranging from many of the students being excited about these experiments and making a connection between the mathematical theory and the experiments to only some of the students seeing a connection between these experiments and quantum technologies. This work can help instructors consider the importance and framing of quantum experiments and raises questions about when and how in the curriculum quantum experiments can be best utilized and how to make related learning goals available to all students.
\end{abstract}

\maketitle

\section{Introduction}\label{sec:intro}

Quantum mechanics, one of the pillars of modern physics, has long been seen as particularly difficult for students to learn \cite{singh2015review,johnston1998student}. This is due, in part, to it being very mathematical and abstract \cite{johnston1998student, johansson2018shut, corsiglia2020characterizing}, counter- or un-intuitive \cite{singh2009cognitive,corsiglia2023intuition}, not seen in the real world \cite{hoehn2017investigating,dreyfus2019splits}, and difficult to visualize \cite{mashhadi1999insights,levrini2013encountering,kohnle2014investigating}. These factors can also lead to some students losing interest in the subject \cite{johansson2018undergraduate}. Nonetheless, demand for a quantum workforce is increasing around the world \cite{fox2020preparing, hughes2022assessing,greinert2022future}, and many new educational programs are being designed \cite{asfaw2022building, perron2021quantum, meyer2022todays,hasanovic2022quantum,plunkett2020survey,aiello2021achieving,kaur2022defining}. Although current work in quantum technologies involves both theoretical and experimental components \cite{fox2020preparing, hughes2022assessing,greinert2022future}, much of the education research so far has focused on the theoretical side \cite{singh2015review, meyer2022todays}. 
There are many open questions about how to best utilize experiments to improve students' quantum education and preparation for the quantum workforce and if experiments can provide students a concrete, non-mathematical approach to the field. 

One benefit of incorporating quantum experiments in undergraduate courses is that students have the chance to observe quantum effects with actual experimental equipment rather than just from a textbook. Many instructors have implemented some variation of a sequence of quantum optics experiments, which we refer to as the ``single-photon experiments,'' into their undergraduate courses \cite{galvez2005interference,beck2012quantum,lukishova2022fifteen,borish2023implementation}. This allows students to work with experiments that demonstrate fundamental quantum phenomena, including ones similar to recent Nobel-prize-winning experiments that laid the foundation for quantum information science \cite{aspect1982experimental,bouwmeester1997experimental,castelvecchi2022spooky}. In previous work, we found that one of the most important learning goals for instructors using the single-photon experiments was for students to ``see'' quantum mechanics in real life. In fact, all of the surveyed instructors ranked this goal as somewhat or very important \cite{borish2023implementation}. Many instructors believe there is a large distinction between students performing quantum experiments and watching videos, demonstrations, or simulations (\emph{vide infra}), yet there is no concrete evidence demonstrating exactly what students uniquely learn from working with quantum experiments.

In this work, we perform a phenomenographic study that investigates how students observe quantum effects in experiments and why it is important for them. We are interested in identifying the variation of possible ways students can experience quantum lab experiments, and whether this differs from the experiences of their instructors. We interview both students and instructors who work with the single-photon experiments in undergraduate courses to answer the following research questions:
\begin{enumerate} [label=RQ\arabic*.]
    \itemsep0em 
    \item How do students think about seeing quantum effects in experiments, and how does that compare with instructors’ ideas?
    \item Do students see quantum effects while working with the single-photon experiments, and what contributes to that?
    \item Do students achieve learning goals related to seeing quantum effects while working with the single-photon experiments?
\end{enumerate}

Here, we present answers to these research questions, by first providing a framework for understanding what it means to students and instructors to see quantum effects and then using those ideas to see how effective the single-photon experiments are at helping students observe quantum effects and achieve related learning goals. We begin with a brief description of prior work studying quantum education in Sec.~\ref{sec:background}. This is followed with details about the interviews and analysis methods in Sec.~\ref{sec:methods}. We then present the results of our study in Sec.~\ref{sec:results}, starting with results for the three research questions and ending with a discussion about different ways experiments can be considered quantum, which spans the research questions. We then present implications for both instructors and researchers in Sec.~\ref{sec:implications} and summarize this work in  Sec.~\ref{sec:conclusions}.

\section{Prior research in quantum mechanics: From student difficulties to lab work}\label{sec:background}

Physics education researchers have studied quantum education for decades, with much of the work focusing on students' conceptual understanding of quantum theory. In this section, we briefly summarize some of this work, beginning with the way quantum mechanics is often perceived as being particularly mathematical and abstract \cite{johnston1998student, johansson2018shut, corsiglia2020characterizing}. This has led to many student difficulties in learning quantum concepts and prompted the creation of new curricula focused on simpler mathematical systems (e.g., two-level quantum systems). To help students better visualize these complex topics, a variety of simulations of quantum phenomena have been created, which have been shown to help students learn concepts \cite{kohnle2015enhancing, malgieri2014teaching, mckagan2008developing, marshman2022quilts,greca2014teaching} and improve student interest in the field \cite{baily2012interpretive,baily2015teaching}. However, they do not afford students the opportunity to see quantum mechanics in physical experiments, an aspect that may help students build a quantum intuition \cite{corsiglia2023intuition}. This body of literature motivates our study where we investigate possible learning gains from students seeing quantum experiments themselves in the context of the single-photon experiments. We end this section with a description of the single-photon experiments, their utilization in courses \cite{borish2023implementation}, and the prior results of their efficacy in specific implementations \cite{galvez2010qubit,pearson2010hands,lukishova2022fifteen}.

\subsection{Studies on conceptual learning in quantum mechanics}

Quantum mechanics has long been considered an abstract and mathematical subject \cite{johnston1998student, johansson2018shut, corsiglia2020characterizing} that is challenging to visualize \cite{mashhadi1999insights, levrini2013encountering}. Some students perceive being good at quantum mechanics as being good at performing calculations, instead of modeling or understanding the world \cite{johansson2018shut}. The way many courses do not explicitly bring up interpretations of quantum mechanics can make it difficult for students to connect the abstract math with their conceptual understanding \cite{greca2014teaching, baily2015teaching}. Some research in introductory courses has suggested that instructors can help students think about physics concepts in terms of their everyday lives \cite{hammer1994epistemological}; however, students do not have everyday experience with quantum systems. Students see little relation between quantum mechanics and the real world \cite{hoehn2017investigating,dreyfus2019splits}, but they may be able to make sense of the new ideas by learning that quantum mechanics is not about memorizing how to perform calculations \cite{dini2016case}. Partly because students rarely encounter quantum mechanics in their everyday lives or see it with their own eyes, many find quantum mechanics to be counter- or un-intuitive \cite{singh2009cognitive, corsiglia2023intuition}.

The abstraction inherent in quantum mechanics and the required mathematical sophistication have contributed to student difficulties in learning quantum mechanics \cite{singh2015review}. The sophisticated mathematics needed to describe quantum systems can increase students' cognitive load \cite{singh2009cognitive}, and students have trouble building mental models of quantum mechanics since they cannot support the models with their own experiences \cite{johnston1998student}. Some students report being discomforted by the concepts and the math-physics connection in quantum courses \cite{corsiglia2020characterizing} and feel like the physics is harder when it is less intuitive \cite{corsiglia2023intuition}. The common emphasis on the math at the expense of the concepts has additionally reduced some students' excitement about quantum physics and caused them to switch into other, clearer areas of physics \cite{johansson2018undergraduate}. 

In part due to the challenging nature of the subject, there has been a long history of understanding student reasoning and difficulties within quantum courses, ranging from high school through graduate education \cite{singh2015review, marshman2015framework, cataloglu2002testing, carr2009graduate, krijtenburg2017insights}. Much of this work has investigated student understanding of specific concepts, such as tunneling \cite{domert2004probability, wittmann2005addressing, mckagan2008deeper}, conductivity \cite{wittmann2002investigating}, quantum measurement \cite{singh2001student, zhu2012improving}, and particle-wave duality \cite{mannila2001building, marshman2017investigating, hoehn2019investigating}. Other work has investigated how student reasoning is connected to various aspects of the instruction, such as terminology \cite{brookes2007using}, notation \cite{gire2015structural}, epistemological framing \cite{bing2012epistemic}, and visualization \cite{cataloglu2002testing}. This body of work has shown that students often struggle to come up with good mental models and therefore solve problems by applying known methods of calculation without having a good conceptual understanding \cite{singh2015review, marshman2015framework}.

To help improve students' conceptual understanding, there has been a push towards new curricula in quantum courses, where some are incorporating earlier discussions about two-level systems \cite{kohnle2013new, malgieri2014teaching,greca2014teaching,emigh2020research}. These systems can be used to describe single-photon interference experiments and entanglement of spin-1/2 particles. They are mathematically simpler than continuous systems, allow students to directly think about quantum systems with no classical analogue, and can lead to discussions about interpretations of quantum mechanics and related quantum information applications \cite{kohnle2013new}. Discussions of quantum optics experiments can additionally provide students the opportunity to learn about photons and their properties through the use of experimental evidence \cite{malgieri2014teaching}. Incorporating discussions about interpretations and quantum optics experiments has been shown to increase student interest and improve their quantum reasoning \cite{baily2012interpretive, baily2015teaching}.

\subsection{Visualizing quantum mechanics}

A complementary approach to improving student understanding of quantum mechanics is through the use of visualizations \cite{kohnle2015enhancing, mckagan2008developing}. Good visual representations can help students construct mental models and therefore learn abstract physics concepts, but some visualizations can also lead to student misconceptions \cite{chen2014how}. 
Even when instructors do not explicitly discuss ways to interpret or visualize quantum phenomena, students can develop their own mental images that are different than those intended by their instructors \cite{baily2015teaching, mashhadi1999insights}. It is therefore important to consider productive ways students can visualize quantum mechanics, especially since some students find visualizations necessary to understand quantum theory beyond the mathematical formalism \cite{levrini2013encountering}. Different visual representations have been shown to influence student learning in the context of the single-photon experiments \cite{kohnle2014investigating}.

One way visualizations of quantum mechanics are incorporated into classrooms is with research-based simulations developed to improve student understanding \cite{kohnle2013new, malgieri2014teaching, mckagan2008developing,kohnle2015enhancing,marshman2022quilts,ahmed2022student}. Simulations allow students to visualize parts of physics they cannot directly observe \cite{mckagan2008developing} in addition to helping  students relate quantum mechanics to reality \cite{mckagan2008deeper}, engage in inquiry driven learning \cite{kohnle2015enhancing}, and build intuition \cite{marshman2022quilts}. 
The interactive component of the simulations more than the visual representation (e.g., screenshots) has been shown to lead to student enjoyment of the activity \cite{kohnle2015enhancing}.
Compared with actual experiments, simulations can reduce the cognitive load for students and allow them to explore a system in a situation where they do not need to worry about breaking equipment \cite{kohnle2015enhancing}. Simulations of the single-photon experiments have been shown to improve student understanding \cite{kohnle2015enhancing, marshman2022quilts}. 

Simulations, however, often do not help students understand how knowledge about quantum mechanics was gained from observations, something that can most easily be done with real experiments. A step in that direction is the use of interactive screen experiments, which are multimedia representations of experiments that allow students to interact with certain experimental settings without needing access to the actual experimental set-up \cite{bronner2009interactive}. Interactive screen experiments of quantum optics experiments have been shown to help students learn quantum concepts that are less influenced by classical physics \cite{bitzenbauer2021effect}. Compared with real experiments, simulations and interactive screen experiments are cheaper, less complicated, and do not require access to an experimental apparatus. There are additionally analogy experiments (experiments demonstrating the idea of quantum  phenomena without utilizing actual single-photon states) that require fewer resources and expertise than the full quantum optics experiments \cite{schneider2002simple,dimitrova2008wave}. We are not aware of prior work investigating if there is any learning about quantum mechanics that students can only achieve by working with a physical experiment utilizing quantum states.

\subsection{The single-photon experiments}

Some institutions with enough resources to support it, have started incorporating quantum optics experiments into their curricula to teach about fundamental quantum effects such as particle-wave duality and entanglement.  These experiments often involve a laser passing through a non-linear crystal in which spontaneous parametric down-conversion takes place. During that process, some of the photons in the laser beam are converted into pairs of lower-energy photons that are entangled in energy and momentum. The resulting photon pairs may then be measured simultaneously either to demonstrate their entanglement or to use as a heralded single photon source. Although there is not an exact set of experiments that falls into this category, there are many related experiments that can be done with similar apparatus, which have been incorporated into undergraduate courses over the past 20 years (see, for example, Refs.~\cite{dehlinger2002entangled,thorn2004observing,galvez2005interference,lukishova2022fifteen,pearson2010hands,beck2012quantum}). We refer to any of these similar experiments as the single-photon experiments. 

The single-photon experiments have become popular in the advanced lab community. They are often taught at the Immersion workshops hosted by the Advanced Laboratory Physics Association \footnote{The Advanced Laboratory Physics Association (ALPhA) is an organization aimed at fostering communication and interaction among advanced laboratory physics instructors at colleges and universities in the United States and the rest of the world. More information can be found at \url{https://advlab.org/}.} where new instructors can learn how to implement these experiments in courses. Instructors are continuing to publish papers about new ways to incorporate extensions or similar experiments in undergraduate instructional labs. In our recent work studying the implementation and goals of these experiments across undergraduate courses, we found that they are primarily used in upper-level quantum or beyond-first-year lab courses, although some instructors are beginning to use them in introductory courses as well \cite{borish2023implementation}. Instructors have a variety of goals for using these experiments including helping students learn about quantum concepts, improve lab skills such as aligning optics, gain interest and motivation, and see quantum mechanics, which is the motivation behind this work \cite{borish2023implementation}. 

Some of the instructors who have published new ways to use the single-photon experiments in their courses have additionally studied the effect these experiments had on their students. These experiments have been shown to improve students' conceptual understanding both self-reported \cite{galvez2010qubit} and through an assessment \cite{lukishova2022fifteen}. They additionally can motivate students to want to better understand the theory \cite{pearson2010hands} and to pursue a career in quantum optics and quantum information \cite{lukishova2022fifteen}. A larger scale study across different implementations has not yet been implemented, so we begin that process here by investigating if and how students meet instructor goals related to seeing quantum effects in experiments.

\section{Methodology} \label{sec:methods}

This study follows a previous survey about instructor usage and goals of the single-photon experiments \cite{borish2023implementation}, and allows us to understand in-depth instructors' and students' ideas related to seeing quantum effects and how the single-photon experiments are an example of that. To investigate these ideas across course contexts, we interviewed instructors and students at many different institutions. We first performed semi-structured interviews of 14 instructors who had set up or utilized the single-photon experiments in their courses, and followed this with interviews of 14 students who had each performed a subset of the experiments in at least one physics course within the previous year. To analyze the interviews, we performed a thematic coding analysis, using the emergent results from the instructor interviews as a basis for our analysis of the student interviews. In this section, we present the details of our methodology including limitations of this kind of study.

\subsection{Participants and courses}

This work studies instructors and students from a range of courses and institutions. Instructors were recruited to participate in the previous survey through the Advanced Labs Physics Association, and all U.S. instructors who completed the survey and agreed to be contacted for future research opportunities were invited to participate in follow-up interviews. At the end of the instructor interviews, we asked if the instructors would be willing to forward a recruitment email from us to the students in their course(s) after the students had finished working with the single-photon experiments. Because many of the courses that use the single-photon experiments are offered only once every year or two, we asked instructors to reach out to students who were currently enrolled in their course(s) or who had taken them earlier that academic year. Presumably due to the small class sizes, we were not able to recruit a sufficient number of students initially, so we carried out a second round of recruitment the following semester, during which we additionally reached out to instructors who had completed our survey but had not participated in a interview and instructors we had recently met in the context of the single-photon experiments (e.g., at conferences, through campus visits, etc.). 

Table~\ref{tab:demographics} shows the self-reported demographics of the interviewed instructors and students. Each participant was given the option to self-report their gender, race, and ethnicity in a free response format at the end of the interviews. We combined the responses at a level that balanced honoring the participants' responses with keeping them from being potentially identified.  All students were assigned pseudonyms without any intentional racial or ethnic significance, and we use these pseudonyms along with the pronouns the students requested. The students were in their second year of study or above. 

\setlength{\tabcolsep}{1.8pt}
\begin{table}
\centering
\caption{Information about the interviewed students and instructors along with their institutions, courses, and experiments performed. The instructors worked at distinct institutions and some discussed more than one course. Some of the students were enrolled in the same courses and institutions as each other, so those are counted multiple times.}
\label{tab:demographics}
\begin{tabular}{>{\quad}lcc}
    \hline
    \hline
    & \makecell{Instructors\\(N=14)} & \makecell{Students\\(N=14)}\\
    \Xhline{2\arrayrulewidth}
    \rowgroup{\textit{Individual}}\\
    Man & 12 & 10 \\
    Woman & 2 & 3 \\
    Non-binary & 0 & 1 \\ 
    White & 14 & 10\\
    Hispanic or Latino & 2 & 1 \\
    Asian & 0 & 3 \\
    \hline
    \rowgroup{\textit{Institution}} \\
    Four-year college & 7 & 3 \\
    Master's degree granting & 4 & 3 \\
    PhD granting & 3 & 8\\
    Hispanic-serving institution & 1 & 0 \\
    \hline
    \rowgroup\textit{{Course}} \\
    Beyond-first-year lab & 12 & 12 \\
    Quantum & 6 & 2 \\
    Introductory & 2 & 0\\
    \hline
    \rowgroup\textit{{Experiments}}\\
    SPDC & 8 & 5 \\
    Existence of a photon & 9 & 3 \\
    \makecell{Single-photon interference \\ and/or quantum eraser} & 11 & 9\\ 
    Bell's inequality & 12 & 9\\
    Other & 5 & 6\\
   \hline
    \hline
\end{tabular}
\end{table}

Information about the courses in which the participants utilized the single-photon experiments, as well as their institutions, are also included in Table~\ref{tab:demographics}. The instructors worked at 14 unique institutions, and some discussed multiple courses at their institution. The students were enrolled at seven distinct institutions with between one and three students from each institution, including both the same and different courses. The majority of the students were enrolled in lab courses for students beyond the first year in their major. 

The students worked with between one and four of the single photon experiments in their courses, with the number of students working on the most common experiments reported in Table~\ref{tab:demographics}. These experiments include setting up the spontaneous parametric down-conversion (SPDC) source, measuring the anti-correlation of single photons sent through a beam splitter (``existence of a photon''), a single-photon interferometer (possibly with a quantum eraser), and violating Bell's inequality. Further descriptions of these experiments can be found in Ref.~\cite{borish2023implementation}. The amount of time spent working on these experiments ranged from a single three-hour lab to labs for the entire semester plus a prior course with the same experiments. Some of the students set up the experiments themselves, whereas others slightly manipulated optics and then took data on experiments set-up by their instructors.

\subsection{Interviews}

The primary goal of the instructor interviews was to understand what instructors meant by the phrase seeing quantum mechanics and why it was important to them. This was a follow-up to a previous survey in which all of the instructors ranked the goal \textit{``Seeing'' quantum mechanics in real life} as somewhat or very important, yet it was not clear how the instructors interpreted this goal \cite{borish2023implementation}. 
The interviews included questions about the courses the instructors taught with the single-photon experiments, the idea of seeing quantum mechanics, and other learning goals for using these experiments. Only the section about seeing quantum mechanics is analyzed in this work since the other parts have been analyzed in prior work \cite{borish2023implementation}. Example questions about seeing quantum mechanics include: 
\begin{itemize}
    \itemsep0em 
    \item What does the term seeing quantum mechanics mean to you?
    \item Do you think students in your course see quantum mechanics while working with the single-photon experiments?
\end{itemize}
Additional questions can be found in the Supplemental Material \cite{SM}. 

The student interview protocol consisted of similar questions to those in the instructor interviews followed by additional, specific questions based off of ideas arising from the instructor interviews. The instructor interviews had been completed, but not yet analyzed, by the start of the student interviews. We additionally changed the wording of the phrase ``seeing quantum mechanics in real life'' to be ``seeing/observing quantum effects in experiments.''  We believed this change better encompassed the way instructors were talking about this idea while eliminating some possible confusion for the students. The student interviews included sections about the idea of seeing quantum effects generally, whether the students observed quantum effects while working the single-photon experiments, and pointed questions related to what instructors had told us they hoped students would achieve while working with the experiments. Example questions include:
\begin{itemize}
    \itemsep0em 
    \item Do you think it's important to see quantum effects in experiments?
    \item What specific parts of the experiment caused you to observe these quantum effects?
    \item Did working with these experiments help you build intuition about quantum effects?
\end{itemize}
The relevant part of the student interview protocol is provided in the Supplemental Material \cite{SM}. 

Both sets of interviews occurred over Zoom, and all participants were compensated for their time. The instructor interviews ranged from 49 to 69 minutes, and the student interviews ranged from 33 to 59 minutes. The instructors were interviewed in the spring of 2022 and the students were interviewed between the spring and fall of 2022 about their experiences with the single-photon experiments in courses between fall 2021 and fall 2022.

\subsection{Analysis}

To analyze the data, we performed thematic coding analyses of the interview transcripts. Our initial coding of the instructor interviews is described in Ref.~\cite{borish2022seeing}, where we found 14 emergent themes related to the idea of seeing quantum mechanics. Many of these themes are interconnected, so we assigned all the codes throughout the entire section of the interview related to seeing quantum mechanics. We chose to focus on existence of these themes instead of the number of instructors assigned each code since we wanted to understand the range of possible ideas. 
 
We used the resulting codes from the instructor interviews as an initial codebook for the student interviews and also allowed for new emergent codes. For our first research question about how students think about seeing quantum effects, we coded for existence in the same way as with the instructor interviews because we again wanted to emphasize the extent of possible student views. For our other two research questions, we chose to present the number of students assigned each code in order to investigate the efficacy of the single-photon experiments. Each research question was focused on a specific part of the interview.

The student codebook was created iteratively, with the codes being discussed by the research team throughout the coding process. After completion of the codebook, we recruited a colleague unfamiliar with this project to perform an interrater reliability (IRR) check on a subset of the student quotes, which achieved 94\% agreement. We chose to present the percent agreement instead of Cohen's kappa because of the low prevalence of some codes across the dataset, which can make Cohen's kappa unreliable \cite{gwet2002kappa}. The IRR process led to the clarification of one code name and a few code definitions.

\subsection{Limitations}

There are three primary limitations to this study. First, as with any study at this level of detail, we had a relatively small sample size. To mitigate this, we tried to reach as wide a range of students as possible. The students in our dataset were enrolled in many different courses and used the experiments in different ways, so we effectively averaged over their experiences. This allowed us to better understand the idea of seeing quantum effects more broadly, but we did not have enough data from each individual course to make claims based on the specific experimental implementations. 

Second, our sample may be biased towards students from more well-resourced institutions who had positive experiences with the single-photon experiments. Since students who had a bad experience or felt like they did not learn much from working with the experiments may have chosen not to participate in the interviews, we may be presenting a more positive outlook on the single-photon experiments than the average student experienced. Additionally, we interviewed only instructors and students who had access to working with these quantum optics experiments, so we do not know how instructors and students who have not worked with these experiments think about these topics. Since the single-photon experiments are expensive, the population of students working with them in courses may not be representative of the overall population of students in upper-level physics courses. 

Finally, we are relying entirely on student self-assessment for all the learning goals discussed in this work, including conceptual learning. It has been shown that students are not always reliable at assessing their own learning \cite{lindsey2015do}. Nonetheless, the focus of this work is on non-conceptual learning gains, for which there are no existing assessments and we must rely on students' self-reporting. One of the goals of this work is to identify the non-conceptual learning goals in the hope that better ways to assess them can be established in the future. 
Another potential concern arising from student self-assessment is the delay between students performing the experiments and the interviews. Due to the timing of this study, some of the students were interviewed months after working with the single-photon experiments and some discussed how they did not remember well all of the experiments they had performed. Also, although this may be true no matter how student learning is assessed, it was difficult to distinguish what students learned from working with the experiments compared with other parts of the course in which the experiments were integrated.

\section{Results} \label{sec:results}

In this section, we present the results of our thematic coding analysis, focusing on the student interviews. This is divided into four sub-sections, the first three of which answer our three research questions: what seeing quantum effects means to students, what contributed to students seeing quantum effects with the single-photon experiments, and whether or not students achieved other learning goals related to seeing quantum effects. The final sub-section discusses an idea that runs throughout these three questions related to how quantum an experiment needs to be. 

\subsection{How do students think about seeing quantum effects?} \label{sec:whatIsSeeingQM}

To answer our first research question, we analyzed the section of the interview where the students discussed seeing quantum effects in experiments generally. Some of the students explained their ideas broadly, while others used concrete examples of experiments, including the single-photon experiments, they had performed in various physics courses. This, along with our prior analysis of the instructor interviews in Ref.~\cite{borish2022seeing}, has led to a set of 14 codes that together describe the range of both instructor and student ideas surrounding seeing quantum effects in experiments. We divided these codes into two categories during the coding process. Our emergent codes are:
\begin{itemize}
    \itemsep0em 
    \setlength{\itemindent}{-1em}
    \itemsep0em 
    \item Seeing quantum effects may include...
    \begin{enumerate} [label = ...]
        \setlength{\itemindent}{-2em}
        \itemsep0em 
        \item experiments described by quantum physics.
        \item seeing experimental results.
        \item clear results that require little interpretation.
        \item interactions with the experiment.
        \item seeing and understanding the experimental apparatus.
        \item understanding the theory behind the experiment.
        \item not literally seeing quantum objects.
    \end{enumerate}
    \item Seeing quantum effects can help students...
    \begin{enumerate} [label = ...]
        \itemsep0em 
        \setlength{\itemindent}{-2em}
        \item believe quantum mechanics describes the physical world.
        \item gain familiarity with quantum mechanics.
        \item improve conceptual understanding.
        \item think about philosophy of quantum mechanics.       
        \item learn about topics of technological and societal importance.
        \item generate excitement and motivation.
        \item make learning quantum seem attainable.
    \end{enumerate}
\end{itemize}
Definitions, explanations, and examples of these codes can be found in the Appendix.

Overall, the students and instructors in our data set talked about seeing quantum effects in similar ways. The emergent codes from the instructor interviews worked well to categorize student ideas, and we only made a few wording changes to the codes presented in Ref.~\cite{borish2022seeing} to better align with the student data. The only new emergent themes appearing in the student interviews were about ways in which experiments could be considered quantum. During the iterative coding process, these were combined with the slightly re-worded code \textit{Experiments described by quantum physics} and are further discussed in Sec.~\ref{sec:whatIsQuantum} since they relate to a broader theme that has shown up in other aspects of this work as well.

However, instructors did discuss some of the ideas related to seeing quantum effects with more nuance than the students. This is not surprising since instructors are more knowledgeable about quantum physics, lab work, and the role physics can play in society. Although some students brought up the technological implications of quantum mechanics in the context of seeing quantum effects, some instructors went further and discussed the broader societal implications (e.g., how the students could ``be stewards'' and explain quantum physics to non-specialists or combat misinformation). Some instructors also considered additional kinds of interactions, such as by discussing decision-making instead of just physical interactions. The only theme that did not appear in the student data related to seeing quantum effects is the code \textit{Think about philosophy of quantum mechanics}. This may be because many instructors, both of the set we interviewed and quantum instructors more broadly, often do not focus on interpretations of quantum mechanics \cite{greca2014teaching, baily2015teaching}.

There is no single definition to students or instructors about what seeing quantum effects means, as evidenced by the many emergent codes. All of the students and instructors were each assigned multiple codes, sometimes even for the same quote, demonstrating how these codes represent many interconnected ideas. Although we asked separate questions to try to distinguish the meaning of seeing quantum effects from its importance, these ideas were mixed together in responses. We suspect this is because these are complicated ideas that instructors and students may not have ever had to clearly define for themselves before. The range of ideas provided us a starting point to look at the prevalence of these ideas occurring in the context of the single-photon experiments.

\subsection{What contributed to students seeing quantum effects with the single-photon experiments?} \label{sec:contributionsToSeeingQM}

Overall, students do think they are observing quantum effects while working with the single-photon experiments. When directly asked about this, 13 out of 14 students responded affirmatively, although the certainty in their responses varied from ``I guess'' to ``Yes, 100\%.'' In order to understand what contributes to students feeling like they have observed quantum effects, we assigned codes similar to the first set in Sec.~\ref{sec:whatIsSeeingQM} to parts of the interview where the students were talking about seeing quantum effects with the single-photon experiments. Table~\ref{tab:whatContributesToSeeing} shows these codes along with example quotes and the number of students assigned each code. All of the students were assigned at least one of the codes.

\begin{table*}[htbp]
\caption{Codes with example quotes describing what contributed to students observing quantum effects while working with the single-photon experiments. The number of students (out of 14) is also included. The * denotes the one code that does not contribute to students feeling like they observed quantum effects, but was often discussed in relation to it.}
  \centering
  \setlength{\tabcolsep}{1.8pt}
    \begin{tabular}{>{\hangindent=1em}>{\raggedright}p{6.4cm} >{\hangindent=1em}>{\raggedright}p{10.4cm} c}     
    \hline
    \hline
    \textbf{Code}& \textbf{Example quote} & \textbf{N}\\
    \Xhline{2\arrayrulewidth}
    Seeing results contributed to students observing quantum effects & ``It was like the, when we were calculating the $g$ whatever values, the second order coherence. Just doing those calculations, cause we set up Mathematica or Python programs to do them, and then just seeing all the values come out to what we would expect them to be, if they were quantum, just like knowing that. And even when we were just looking at the detectors, like the raw sort of numbers coming out of them, we could just tell that if it was classical these would be a lot higher. So, I guess the detection itself.''& 13\\ 
    Experiments described by quantum mechanics contributed to students observing quantum effects & ``I certainly saw the quantities that... we theoretically showed were tests of quantum mechanics, tested quantum predictions. And we showed that they did in fact either break the classical prediction or confirm the quantum prediction. And so in that sense, yes, I guess I did see quantum effects.'' & 11\\ 
    Understanding the underlying theory contributed to students observing quantum effects & ``It was the whole thing... It is easier to see the interference pattern emerging. So I think, eh, it's a wave. But then when you'll see that, how the photon's being produced, when you actually carry on the calculation to see how many photons should be in the interferometer, that's when it clicks that okay, this cannot be explained by some easier explanation.''& 9\\ 
    Understanding the apparatus contributed to students observing quantum effects & ``Both of the setups had beam splitters, they had like polarizing plates and films that we were able to adjust and move. And so, especially for the wave nature that was really helpful for me to sort of conceptualize like okay, it goes through this and what makes it through... which part goes which way, and how does it recombine and where does it go... I think that that was a really sort of helpful way to sort of represent the phenomenon that we were talking about.'' & 8 \\
    Interactions with the experiment contributed to students observing quantum effects &  ``Once the whole setup was done, you can just shift any one of elements. You can adjust the polarizers and you can see the actual shift in your numbers... on the screen as it's running. And so just being able to see that the effect that rotating the polarizers has as well on your computer screen, really, was another part that really just like, you could see the experiment working.'' & 7\\ 
    Clear results requiring little interpretation contributed to students observing quantum effects & ``So one part that really kind of affirmed that I was seeing it...  we were turning the phase plates, just kind of changing the angles and stuff like that. And so you could clearly see that when one was vertically polarized, and one was horizontally polarized, then there would be no coincidences... there was another portion where we just kind of kept changing the angle and we'd kind of make a graph out of the number of coincidences. And you could clearly see the sine wave, which was what was supposed to happen. And so that was more proof that it actually happened.'' &5\\ 
    Acknowledge cannot see quantum objects & ``And it's hard with quantum too, because it's like I want to see it play out, but these particles are too small, and I can't really see them.'' & 4* \\ 
    \hline
    \hline
  \end{tabular}
  \label{tab:whatContributesToSeeing}
\end{table*}

\subsubsection{Varied combinations of codes contributed to students observing quantum effects} \label{sec:contributionsToSeeingQM-differentCombinations}

Multiple aspects of working with the single-photon experiments contributed to students feeling like they observed quantum effects. On average, each student was assigned more than three of the codes in Table~\ref{tab:whatContributesToSeeing}, although the exact combination varied by student. The most prevalent code was \textit{Seeing results contributed to students observing quantum effects}, which was often assigned at the same time as other codes. For example, students discussed how the results  matched quantum and not classical predictions, how seeing the results let them know if the results were particularly clear, how seeing the results with the theory in their head helped them understand what was happening, how they interacted with the experiment and then looked at the results, and how they understood the results though understanding the different parts of the apparatus. Many of the quotes in Table~\ref{tab:whatContributesToSeeing} include both the listed code, as well as a part about the students looking at the resulting data or graphs. Possibly because of the different ways students needed to engage with the experiments in order to observe quantum effects, many students discussed how they observed quantum effects more in some experiments than others within the same course.

Not all aspects were useful to all students. For example, Jaime explained how he understood each component of the apparatus, yet that ``is not what makes my mind click OK, this is quantum or classical. This is just an instrument for me to get data.'' Additionally, not being able to physically see the photons mattered more to some students than others. Hayden, the only student for whom it was not clear that he thought he had observed quantum effects while working with the single-photon experiments, explained: 
\begin{quote}
\textit{``I know that we did. Because obviously I saw the Bell inequality being violated. I saw the coincidence counts of the entangled photons and everything. But I guess it was just a little harder to... actually know that it was quantum effects. Because with normal physics experiments, you can see everything that's happening... But with quantum, like everything's so small... you can't see anything. And so I saw all these numbers and everything, but... they could have been... random computer generated for all I knew... So I did witness it, but I didn't like really} witness \textit{it.''}
\end{quote}
On the other hand, the other three students who acknowledged they could not see the down-converted photons felt that they were observing quantum effects through seeing the results and understanding how each part of the apparatus affected the photons.

\subsubsection{Students focused on different aspects of the results when determining what was clear}

Some of the students talked about not only seeing the results, but having some aspect of the results be particularly clear as an important part of seeing quantum effects. There were two distinct ways students talked about the clarity of experimental results. First, similar to the quote in Table~\ref{tab:whatContributesToSeeing}, some students discussed how it was obvious that the results they were seeing lined up with the experiment. This was often brought up when talking about rotating a polarizer and seeing the counts of photons increase and decrease. Other students instead discussed how the prediction for whether the results were quantum or classical was clear. For example, when comparing different experiments he performed in his course, Casey said:
\begin{quote}
    \textit{``I think it's because those experiments were far more formulated in terms of a classical versus quantum prediction... They both were sort of similarly formatted in the sense that... here's this quantity that effectively, that allows us to test, directly to test a classical versus quantum prediction, and now let's measure it. And I think that's sort of a straightforward formulation. I think being able to formulate it in a straightforward manner, made it sort of easier to see.''}
\end{quote}

These two kinds of clarity may not always be compatible, as was evident with the different ways students discussed the Bell's inequality experiment. Students who fell into the first category (those who wanted it to be clear how the results lined up with the experiment) expressed that simpler experiments were clearer to them than Bell's inequality. When asked to compare different quantum effects he saw in his course, Briar said: ``I mean, obviously the Bell inequality was really important because it sort of proved what we were doing, but the Malus' law things, to me, sort of gave a more tangible example of what was happening.'' By Malus' law, he is referring to rotating a polarizer and detecting how the change in counts varied based on the polarizer angle. On the other hand, for students focused on a prediction that clearly distinguishes between quantum and classical outcomes, Bell's inequality may be very clear. When discussing how he observed quantum effects while working with this experiment, Nicky said: ``And the number that pops up on the screen after we take the coincidence data is 2.8 or whatever, or 2.3... that clearly goes against the assumptions of local realism.''

\subsubsection{The depth and timing of the requisite theory was varied}

Another common code was related to students understanding the theory underlying the experiments. Students had varied responses about how much of the theory they needed to know in order to observe quantum effects, ranging from all of the details to just parts of it. For example, Jaime discussed how in experiments that were not obvious, it was necessary to ``fully understand the quantum theory in order to think yeah this is quantum not classical.'' On the other hand, Casey said that he ``didn't quite fully get the derivations'' for the two experiments that he felt were ``satisfying in a sense that I felt like I observed a truly quantum mechanical effect.'' These were also the two experiments he remembered best after his course was over. Nonetheless, of those two experiments, the one that was more satisfying for Casey was the one that was easier to understand. 

The timing of when the students needed to understand the theory also varied by student. For example, Frankie, when asked if it mattered to her whether she understood the concepts before or after taking the measurement, said:
\begin{quote}
    \textit{``Oh definitely before... You understand the concepts, you set up your equations, and then after, you compare the results to them. So, it's good to know what you're doing or why you're looking for what you're looking for.''}
\end{quote}
For other students, it was sufficient to understand the theory after performing the experiment, such as while writing up the accompanying lab report. When asked if he needed to fully understand the concepts in order to observe quantum effects, Hayden said:
\begin{quote}
     \textit{``I don't think you need to understand all the concepts to feel like you're witnessing quantum effects, because, I remember, while I was doing the experiment I didn't fully understand the equations that went into it. I didn't really understand what a Bell inequality was. I just followed the directions, and I just kind of like saw everything happening as it was supposed to... Later on, when I actually sat down to write the paper and actually learned all the concepts that went into it, I was like oh wait, this is why this happened.''} 
\end{quote}
Nicky, who also talked about only feeling like he had observed quantum effects when he had gone through the derivations after performing the Bell's inequality experiment, acknowledged that he would have found it less frustrating and ``appreciated it more'' if he understood the reasoning behind the procedure while working with the experiment.

\subsection{Did students achieve learning goals related to seeing quantum effects while working with the single-photon experiments?} \label{sec:learningGoals}

Independent of whether or not students are observing quantum effects by their own definition, they may still accomplish other learning goals related to seeing quantum effects set by their instructors. To answer our third research question, whether students report achieving some of these related learning goals by working with the single-photon experiments, we looked at student responses to questions based on the second set of codes related to seeing quantum effects that were discussed in Sec.~\ref{sec:whatIsSeeingQM}.

\subsubsection{Students achieved many learning goals}

Table~\ref{tab:goalCodes} shows the emergent codes from this analysis, grouped into categories, and the number of students assigned each code. Almost all of these codes came from specific questions about those topics (instead of about seeing quantum effects more generally), so most students were assigned at least one of the codes for each question. However, for each set of codes, the numbers do not add up to 14 students for several reasons. First, there were some instances where students ended up talking about related topics without directly answering the interview questions, so we could not classify their responses. Second, some students discussed seemingly incompatible ideas at different parts in the interview, so they may have been assigned two ostensibly contradictory codes. This often occurred when students discussed either two different experiments or a specific part of one experiment compared with their experience overall. Note also that there are varying levels of agreement for the students assigned to each code; we were not able to capture all of the nuances. 

\begin{table*}[htbp]
\caption{Codes related to goals instructors have for their students that are connected with seeing quantum effects (the second set of codes in Sec.~\ref{sec:whatIsSeeingQM}), and the number of students (out of 14) assigned each one.}
  \centering
  \setlength{\tabcolsep}{1.8pt}
    \begin{tabular}{>{\hangindent=1.5em}>{\raggedright}p{13cm} c}   
    \hline
    \hline
    \textbf{Code}& \textbf{N}\\
    \Xhline{2\arrayrulewidth}
    \textit{Beliefs about quantum mechanics describing the physical world} &\\
    \hspace{1.5em} Confirmed students' beliefs in the validity of quantum mechanics & 8 \\
    \hspace{1.5em} Did not change students' beliefs about the validity of quantum mechanics & 5\\
    \hspace{1.5em} Had never doubted the validity of quantum mechanics & 10\\
    \hspace{1.5em} Made math--experiment connection  & 11 \\
    \hspace{1.5em} Did not make math--experiment connection  & 4\\
    \textit{Familiarity with quantum mechanics} &\\
    \hspace{1.5em} Made seem more weird or mysterious & 3\\
    \hspace{1.5em} Made seem less weird or mysterious & 5 \\
    \hspace{1.5em} Did not change weirdness or mysteriousness & 7\\
    \hspace{1.5em} Gained intuition or familiarity with quantum mechanics & 11\\
    \hspace{1.5em} Was already familiar with quantum mechanics & 4\\
    \textit{Excitement and motivation} &\\
    \hspace{1.5em} At least part of experiment was exciting & 13 \\
    \hspace{1.5em} Did not find experiment exciting & 1 \\
    \hspace{1.5em} Alignment diminished excitement & 3 \\
    \textit{Conceptual understanding} & \\
    \hspace{1.5em} Learned about quantum concepts & 10 \\
    \hspace{1.5em} Learned concepts about experimental apparatus & 11 \\
    \hspace{1.5em} Learned concepts about quantum statistics and uncertainty & 3\\    
    \hspace{1.5em} Learned concepts (other or unspecified) & 7 \\
    \hspace{1.5em} Still do not understand some part of theory of quantum effects & 3\\
    \hspace{1.5em} Still do not understand some part of apparatus & 7 \\
    \textit{Technological importance} &\\
    \hspace{1.5em} See at least some connection between experiments and quantum technologies & 8\\
    \hspace{1.5em} Do not see connection between experiments and quantum technologies & 2\\
    \hspace{1.5em} Do not know about quantum technologies & 5\\
    \textit{Other} &\\
    \hspace{1.5em}  Feel more capable of understanding quantum mechanics & 2 \\
    \hspace{1.5em} Thought about philosophy of quantum mechanics & 1\\
    \hspace{1.5em} Thought about what is needed for quantum results to be valid & 3\\
    \hline
    \hline
  \end{tabular}
  \label{tab:goalCodes}
\end{table*}

Overall, students achieved many of the learning goals instructors had related to the idea of seeing quantum mechanics, although there was variation by student of exactly which goals and the degree to which they were accomplished. All of the students were assigned at least one code related to a positive outcome, and the majority of the interviewed students reported achieving many of these positive outcomes. These include making a math--experiment connection, gaining intuition or familiarity with quantum mechanics, confirming a belief about the validity of quantum mechanics, being excited by at least part of the experiment, improving their understanding about quantum concepts and the experimental apparatus, and seeing at least some connection between the experiments and quantum technologies. On the other hand, there are some negated versions of these codes as well, showing that not all students accomplished the different goals.

\subsubsection{Improved conceptual learning and math-experiment connection}

Although the majority of beyond-first-year lab courses, where the single-photon experiments are most commonly implemented, focus on developing lab skills over reinforcing concepts  \cite{holmes2020investigating}, instructors using the single-photon experiments often have learning goals related to student conceptual learning about topics such as particle-wave duality, entanglement, and quantum states \cite{borish2023implementation}.  Many of the students in our dataset believed that working with the single-photon experiments helped them learn about quantum concepts and make a connection between the math and the physics. 

At least 10 of the students in our dataset reported learning about quantum concepts. This is a lower bound because other students discussed improving their learning in some capacity, without specifying whether it was about quantum concepts or some part of the apparatus. The only students who mentioned not learning about quantum concepts while working with the single-photon experiments did so when discussing how they understood the concepts well already. The students in our study had a large range of levels of conceptual knowledge about quantum mechanics before working with the single-photon experiments; some students had not previously taken a quantum class, while others had taken several or had studied the subject on their own. Some of the students coming in with a solid grasp on the concepts still thought they improved their understanding while working with the single-photon experiments. For example, when asked if these experiments helped him better understand quantum concepts, Indigo said:
\begin{quote}
    \textit{``Yeah, I think it really did. I actually think that even though before the experiment, I would have thought, yeah I understand this concept, I feel like I had moments in the experiment where I had somewhat unexpected results, and there was some nuance to the experiment which I didn't know, which led to results that we didn't expect and didn't get us the right results. And because of those nuances, I had to better understand what was happening conceptually on the quantum level to fix those results... That's a conceptual understanding that I don't think I could have gotten by just learning about it. I feel like I had to go through that process of having to solve the problem in the experiment to understand those kind of things.''}
\end{quote}

In addition to improving their conceptual understanding, many students felt that they were able to make the connection between the math they had studied in prior courses or read about in relation to the experiment and the experimental results they found. Eleven of the students made a math--physics connection for at least one part of the experiment. For example, Dana talked about how he had learned about qubits mathematically prior to working with the single-photon experiments, yet while manipulating the optics of the experiment, he was ``able to put an experimental setup to the theory.'' This taught him how he could experimentally test out the math. Nicky discussed how understanding the connection between the procedure and the theory helped him better understand the concept of entanglement: 
\begin{quote}
    \textit{``I think everyone has this kind of sense of what we mean when we say quantum entanglement. It's kind of like, it's just fuzzy. But I think this, and like the process, the experiment kind of, just kind of fixes it and connects it to... the actual physics.''}
\end{quote}
Just as with seeing quantum effects generally, various aspects of the experiment helped students make this connection, including performing the data analysis, manipulating the experiment and seeing the result, and evaluating if the experimental result matched the student's expectation.

There were only three students for whom working with the single-photon experiments did not help them make this connection, with one additional student who was not able to make the connection for one specific part of the experiment. These students cited several reasons for not making the connection including that ``there's always a little bit of a disconnect,'' the experiment ``didn't really require that much mathematics at all,'' or the student did not ``have the math to back it up.''

\subsubsection{Quantum was not surprising for students, but experiments still provided confirmation} \label{sec:notJarring}

Quantum mechanics did not seem to be particularly jarring or surprising to the students, but they still obtained benefits from working with the experiments. For example, the majority of the students had never doubted the validity of quantum mechanics, yet many of them discussed how seeing the experiment still confirmed their belief in the field. When asked if his views of the validity of quantum mechanics changed after working with these experiments, Casey said ``I didn't disbelieve it before, but I definitely believe it more now.'' Frankie talked about how her excitement for working with the single-photon experiment came about from wanting to be the one that proved that photons existed:
\begin{quote}
    \textit{``...it was the first one I chose, because I saw it said proving photons exist, and I was like I want to do that, I need to do that. I need to know, and I need to do it myself, just to like make sure.''}
\end{quote}
Confirming previously known results was important for the students when working with concepts that were not intuitive or easy to understand.

The single-photon experiments helped students become more familiar with quantum mechanics in other ways as well. Prior to working with the experiments, most of the students thought quantum mechanics was weird or mysterious, and working with the experiments helped five of the students think of it as less weird or mysterious than before. For Logan, this was because working with the experiment made it so quantum ``seem[ed] like a more everyday thing.'' For other students, working with the experiment did not change their views on the weirdness of quantum mechanics, or even made them think of the subject as more weird or mysterious due to learning about concepts they had not previously known existed. Morgan, when discussing how she did not have the math to fully understand how the down-conversion crystal worked said ``And I'm worried that's why it's mysterious. Just like any unexplained effect is automatically magic.'' When asked if working with these experiments had helped them gain intuition about quantum effects, four students said they were already familiar with quantum ideas through self-study or popular media, yet two of those students still talked about gaining additional intuition by working with these experiments.

One student also discussed how seeing an experimental realization made him realize that quantum experiments were not as difficult as he expected based on the theoretical thought experiments discussed in his course. When discussing if his views about whether or not quantum mechanics is weird or mysterious changed after working with the experiment, Greer said: 
\begin{quote}
    \textit{``Having worked with experimental setup and stuff, it's easier to work with on an experimental level than I kind of would have expected doing the math... when we go back to our quantum mechanics class, we're always talking about... you have an oven that's shooting electrons out. And you're like what's an electron oven? I don't know what that means... Our experimental setup just kind of taught me that it's easier to work with than I kind of initially expected it to be in class... There's easier ways to find the effects and measure the effects than the theoretical perfect setups we talked about in class.''}
\end{quote}

\subsubsection{Students were generally excited}

Almost all of the students were excited to work with the single-photon experiments during at least part of the process. When asked if he found these experiments to be exciting or motivating, Casey discussed how these experiments were ``far more interesting than the labs that I've done previously'' because of both cool equipment and the way the experiments measure ``more sophisticated predictions.'' He continued to explain, ``I mean showing that photons have to exist is a pretty big deal as opposed to just finding the gain of a particular filter.'' Other students also talked about their excitement about the concepts covered in the single-photon experiments. When asked what he found exciting about the Bell's inequality experiment, Briar talked about how cool the experiment's goal was: 
\begin{quote}
    \textit{``I mean it's sort of those things where... it starts off as super buzzwordy, where you're like oh, you know this experiment breaks reality and defies our assumptions about the way the world works. And you know it sounds super cool to start with, and then you find out that that's true. That's not just like sort of a sensationalized advertisement. That's actually what we're doing with the experiment is challenging our assumptions about reality.''}
\end{quote}
Whether or not this excitement led to students being motivated to pursue physics more broadly is still an open question.

Although most students were excited about these experiments, one student was never excited about them and three students mentioned how the large amount of time spent on optical alignment detracted from their excitement. Some students liked the alignment aspect of the single-photon experiments, but for others it caused frustration, even if they may have been excited about the results. Nicky discussed how he had been excited going in to the experiment, but how ``start[ing] from scratch after like not having the lasers and stuff aligned... kind of diminished my interest in it.'' This emphasizes the concern many instructors had about the challenge of figuring out the optimal amount of alignment students should perform \cite{borish2023implementation}.

\subsubsection{Connection to quantum technologies}

Some of the students identified a connection between the single-photon experiments and quantum technologies, such as quantum computing and quantum cryptography, while others did not. This mix of student responses is not too surprising since this was important to some instructors, but not others. Although eight of the students claimed they saw at least some connection between the single-photon experiments and quantum technologies, not all of the connections the students discussed were necessarily accurate. Nonetheless, we assigned codes based on the students' assessments of whether or not a connection existed because that could motivate the students to pursue a career in quantum information science and technology.

The connections the students saw between the single-photon experiments and quantum technologies ranged from seeing basic quantum principles in action to ideas of how these could be applied. Some students discussed thinking about light as photons or seeing a superposition or entanglement in practice. Other students were more explicit about the perceived connections, such as how entangled photons can ```transfer' information across time and space'' as a way that quantum computers work or how ``erasing data, but then kind of bringing it back'' could be useful for quantum cryptography. Several of the students acknowledged that photons are not the dominate platform currently being used for quantum computing, but still saw the experiments as at least somewhat related in that they allowed experimenters to ``test ideas and to test better solutions,'' or to ``measure these effects at this small time scale.''  

There were a few reasons students did not see a connection between the single-photon experiments and quantum technologies. This occurred because the course focused on fundamental tests of quantum  mechanics instead of applications, because the students did not know enough about quantum technologies, or because the students just did not see a connection. Five of the students reported not knowing enough about quantum technologies to say whether or not there was a connection.

Even if students did not see a connection to quantum technologies, they may have learned about experimental processes that could help them pursue a related job. All of the students reported learning at least one method of experimentally creating quantum superposition or quantum entanglement, a key resource for the current generation of quantum technologies. The students had a range of confidence in their answers. Some students could not explain anything beyond shining a laser through a crystal whereas other students generalized slightly more. Nicky discussed how he did not know any other methods to generate entanglement, but that he would need ``some physical process that kind of imbues chance into the experiment.'' Even the students who had already learned in prior courses how to theoretically generate entangled particles learned something additional from working the experiment. When discussing how his explanation about experimentally creating quantum entanglement would have changed from before working with the single-photon experiments to after, Kai said: ``I don't think I would have been necessarily wrong, but I think my answer... might be longer now.''

\subsection{What is quantum about experiments?} \label{sec:whatIsQuantum}

One of the codes in Sec.~\ref{sec:whatIsSeeingQM}, \textit{Experiments described by quantum physics}, is seemly obvious, but also very important for whether or not students and instructors feel like they are seeing quantum effects. Although we did not explicitly ask about this in the context of observing quantum effects, the idea of how quantum mechanics manifests in an experiment or how ``quantum'' an experiment needs to be came up in several ways throughout these interviews. Here we discuss three aspects of this idea: (1) a comparison of ways quantum effects can be exhibited in different types of experiments, (2) a comparison of some students' experiences with ``single-photon'' sources of various degrees of authenticity (e.g., heralded single photons versus an attenuated laser), and (3) a discussion of student responses when asked what is quantum about the single-photon experiments.  

\subsubsection{Ways quantum mechanics manifests in experiments} \label{sec:quantum-waysManifests}

There are a variety of quantum effects one can observe in an experiment. Some students discussed having performed prior labs with equipment that was based on quantum mechanics (e.g., lasers and detectors), but explained how those were different than the single-photon experiments. Indigo pointed out how when the equipment was based on quantum mechanics, ``the quantum effects were kind of like side parts of the whole experiment; they weren't too important.'' That was in contrast to the single-photon experiment where ``the entire part of the experiment was about these quantum effects.'' This also came about when Dana discussed the results from experiments in an electricity and magnetism lab course he had previously taken; instead of ``measuring the quantum effects,'' he measured quantities such as current and magnetism so ``it doesn't really count.'' Alex discussed how quantum effects usually showed up as ``errors because we aren't accounting for this quantum mechanical effect.'' Although Indigo and Dana did not think experiments with quantum effects integrated into them were as useful for seeing quantum effects as experiments where it was the focus, Alex discussed how these were ``two sides of the same coin,'' and went on to explain why she thinks both are important: 
\begin{quote}
    \textit{``I think it's good to see examples of both. To see if we're sort of isolating a quantum effect, how do we examine it and how do we study it and what can we learn from it. But at the same time, seeing that it is everywhere and it's integrated into a lot of the other things that we do.''}
\end{quote}

Students and instructors also compared the way quantum mechanics showed up in the single-photon experiments with other experiments that also explicitly utilized quantum phenomena. When comparing the single-photon experiments with a lab using a scanning tunneling microscope, Alex said: 
\begin{quote}
    \textit{``The entirety of the experiment was more built around sort of coming to a conclusion about quantum mechanics as opposed to we're using a quantum phenomenon to do something else, where you still see it at play, but it's not necessarily like explicitly focused on the quantum stuff that we're talking about.''}
\end{quote} 
One of the instructors compared the single-photon experiments with experiments demonstrating nuclear decay, by saying that the students ``don't really get into the nuclear decay models, right, it's really phenomenological''; however, with the single-photon experiments,
\begin{quote}
    \textit{``...you can actually do the analysis. You can write out and solve for the interference pattern. And so I think that's one of the values of it is it's a very tangible way to sort of use quantum mechanics... [In] these experiments, you can write down the wave function essentially and follow it through. So I think that's maybe what's a little bit different than a lot of the other experiments... you have to use the quantum formalism, the math, a little bit to solve, to analyze the data. Or to explain the data... It really feels like they're having to use the quantum mechanics, not just understand the phenomenon.''}
\end{quote}
There are some ways quantum mechanics exhibits differently in the single-photon experiments than other quantum experiments, and this made some, but not all, students feel like they could more easily see quantum effects with these experiments.

\subsubsection{Different kinds of ``single-photon'' sources}

Although the single-photon experiments we have been focusing on use light in a single-photon state (when the other photon in the pair is detected), some similar experiments instead use attenuated lasers. These put neutral density (ND) filters in front of a laser until on average there is less than one photon in the experiment at a time; however, the state of the light is a coherent state and not a Fock state. This distinction is more important to some of the instructors and students than others. One instructor explained how these are entirely different: 
\begin{quote}
    \textit{``...there are experiments out there purporting to be single-photon experiments that are just a HeNe laser with an ND filter. And I don't think that's right. I think you can say that on average you've got less than one photon in there, but if it's a coherent state with $\bar{n}$ less than one, it's still a coherent state...  But I think that's too much for me to get at my [students] with.''}
\end{quote}
On the other hand, another instructor discussed how an interferometer with an attenuated laser was ``a good example of seeing quantum mechanics,'' but still went on to claim that the single-photon experiments ``are even better'' because they are ``clearer'' and ``more irrefutable.'' This instructor also  acknowledged that the attenuated laser not being an actual single photon source ``doesn't occur to [the students].'' 

Some of the students in this dataset worked with single-photon interferometers where the ``single photons'' came from highly attenuated lasers instead of heralded single-photon sources. We did not ask about this distinction since our focus was on the experiments using heralded single-photon sources, so we cannot make any definitive claims about the efficacy of the two options. However, even the students who used attenuated lasers talked about observing quantum effects with them. One student who had the opportunity to work with both of these light sources discussed the different benefits from working with each.

Setting up an interferometer with an attenuated laser is easier and cheaper than doing so with heralded photons, so future work could investigate which learning goals can be achieved with each option.

\subsubsection{What students think is quantum about the single-photon experiments}

When directly asked what is quantum about the single-photon experiments, students gave a range of responses, some more and some less expert-like. Many of the students contrasted different behaviors they would expect for phenomena explained by quantum versus classical models, in a similar manner to how several of the instructors discussed seeing quantum mechanics. A few students, however, mentioned that these experiments were quantum because they dealt with small particles, without any further explanation. When asked if he understood what was quantum about the Bell's inequality experiment and how that differs from classical ideas, Briar said:
\begin{quote}
    \textit{``I think so. I mean from from what I understand, the word quantum sort of means like the smallest possible unit of something. And so it's like if you're working with a photon, that's like the smallest possible unit of light.''}
\end{quote}
When later asked if he had thought about how Bell's inequality is quantum aside from a violation of it being demonstrated with small particles, Briar said:
\begin{quote}
    \textit{``I don't know. I haven't really thought about the inequality itself, other than just how we got it and what it means about sort of the nature of reality... How can an equation itself be like quantum?''} 
\end{quote}
Not all students may be thinking about why these experiments are quantum in the same way as instructors.

\section{Implications and future work} \label{sec:implications}

In this section we synthesize the results and present a few take-away points for both instructors utilizing these experiments in courses and researchers interested in continuing to study the ideas presented in this work. Instructors can use these results when thinking about how to help their students feel that they are observing quantum effects, ways to frame their experiments to help students achieve related learning goals, and how best to assess student learning with these experiments. Researchers can follow-up on this work to further investigate best practices for using these experiments in different contexts across the physics curriculum, the uniqueness or lack-there-of about quantum for student outcomes related to observing experimental effects, and strategies for achieving learning goals related to quantum experiments with limited resources.

\subsection{Implications for instruction}

The single-photon experiments were effective at causing the students in our study to feel like they observed quantum effects. Many students felt like they did so by seeing experimental results in combination with interacting with the experiment, understanding the experimental apparatus, or understanding the theory behind the experiment. Whether or not students thought they were observing quantum effects was additionally affected by how the experiment was framed in terms of quantum versus classical models and how clear the results were without additional interpretation. Because not all students feel that they observe quantum effects by taking the same actions, instructors may consider providing opportunities for all of these actions within their courses. Instructors should acknowledge that what is needed for students to feel like they are seeing quantum effects may be somewhat different than for the instructors themselves. 

There are some goals related to seeing quantum effects that students are likely to achieve while working with the single-photon experiments no matter the differences in instruction, while other goals may depend more on how instructors frame the experiments. For example, if instructors' goals are to help students believe that quantum mechanics describes the physical world or have some excitement about the experiments, instructors may not need to explicitly focus on those goals. Almost all of the students in our dataset achieved some positive outcome related to both of those, although the students we report on did self-select into this study and thus are likely to have had positive experiences. However, if instructors goals are more specific, such as thinking about the philosophy and interpretations of quantum mechanics or learning about the current technological and societal importance of quantum experiments, instructors may need to intentionally build these ideas into their courses. Students are less likely to consider these more nuanced ideas on their own. 

Throughout these interviews, we heard some student comments about quantum mechanics in general or these experiments in particular that were not always completely correct, although the students were not aware of it. Since we were not focusing on students' understanding of quantum concepts, we did not probe this deeply. We do not know whether this came about from imprecise use of terminology; students' experiences with the experiments; or outside sources, such as popular media, where quantum ideas to various degrees of accuracy are becoming more prevalent \cite{meyer2023media}. If instructors want to ensure their students have successfully grasped all the complex concepts exhibited in the single-photon experiments, they could consider assigning an assessment specifically geared towards those ideas or investigate student learning through some format other than student self-assessment.

\subsection{Implications for future research}

There are many remaining open questions about best practices for various methods of incorporating the single-photon experiments into courses. In this study, we were able to show that certain ways of working with the experiments led to students being more likely to observe quantum effects, but we were not able to distinguish student responses by course type or the way the experiments were implemented. These experiments are incorporated into quantum courses with large lecture components, beyond-first-year lab courses, and even some introductory courses, and the way they are integrated into the courses can depend on the context and the course itself \cite{borish2023implementation}. Future work could investigate which of the learning goals related to seeing quantum effects students learn from different implementations of these experiments, what specifically contributes to those learning goals in each context, and how to most effectively use the experiments in each. A large-scale survey could be implemented to obtain these data from a wider range of students.

Instructors and students both discussed the importance of seeing quantum effects in a lab, with some thinking it is more important than seeing other areas of physics and others thinking it is equally as important as seeing other areas of physics. Most of the codes in Sec.~\ref{sec:whatIsSeeingQM} related to how seeing quantum effects can help students could be seen as examples of the purpose of physics experiments more generally \cite{hu2015framework}. However, some of our codes may be more relevant for quantum mechanics than other areas of physics due to the way the field is often perceived and the public awareness of quantum information science and technology. Further work could examine the degree to which the importance of seeing experimental effects is unique to quantum or whether it is similarly important for other areas of upper-level physics. 

This work implies that students do think they are learning and obtaining benefits from seeing quantum effects in experiments; however, many research-level and instructional quantum experiments are more expensive than many institutions can afford. The single-photon experiments are a way to bring quantum experiments to students at undergraduate focused institutions without expensive research labs, yet they still cost tens of thousands of dollars. Future work should investigate cheaper ways to achieve some of the same learning goals including addressing questions such as: What other kinds of quantum experiments not considered here could be cheaper and still lead to students feeling like they observe quantum effects? Which learning goals can be accomplished with an attenuated laser instead of a heralded single-photon source? Which learning goals can be accomplished with cloud based quantum experiments \cite{north2023albert, IBMquantumExperience, IonQquantumCloud}?

These questions are especially important as a few students in our study discussed unprompted how these experiments made them feel more capable of understanding quantum mechanics instead of being intimidated by it. As physics educators are trying to make the field more approachable to a wide variety of students, more work needs to be done to investigate if seeing quantum effects in experiments can help with that, and if so how to make these or comparable experiments available to students at institutions with all levels of resources.

In this discussion, we have been focusing on student learning that can be achieved by working with experiments; however some possible learning goals (such as improving conceptual understanding) have already been demonstrated in free platforms such as simulations \cite{kohnle2015enhancing, marshman2022quilts} and interactive screen experiments \cite{bitzenbauer2021effect}. More investigation is necessary to understand if there are certain concepts that are easier learned about with one platform or another, and to directly compare the benefits to students of the different platforms for the overlapping learning goals.

\section{Conclusion} \label{sec:conclusions}

The idea of students observing quantum effects in real experiments was important to at least some degree to all interviewed students and instructors. For some, it was important because quantum mechanics is a pillar of modern physics, while for others, it is particularly important due to the way quantum is thought of as abstract, mathematical, and hard to see. The single-photon experiments, in part due to the way they demonstrate basic quantum phenomena, overall helped students feel that they were observing quantum effects and achieve other related learning goals, although there was variation across students. This work provides suggestions for instructors about different parts of the experimental process that may help students realize they are observing quantum effects. It additionally raises new research questions about what concepts students are successfully learning with these experiments, the degree to which this is unique to quantum, and how learning goals related to seeing quantum effects can be achieved with fewer resources.

\begin{acknowledgements}
We thank the interviewed students and instructors for participating in this study, Kristin Oliver for performing the IRR check, and the rest of the CU PER group for useful conversations and feedback. This work is supported by NSF Grant PHY 1734006 and NSF QLCI Award OMA 2016244.
\end{acknowledgements}

\appendix*

\section{Definitions of codes about what it means to see quantum effects} \label{app:definitions}

In this appendix, we present more details about what seeing quantum mechanics or quantum effects means to both instructors and students. All of the codes discussed in Sec.~\ref{sec:whatIsSeeingQM} are presented with their definitions in Table~\ref{tab:seeingQMcodeDefinitions} with the following sections discussing each one in detail including example quotes. We focus on student quotes when the student and instructor ideas lined up, and additionally include instructor quotes when they provided more nuance. Other instructor quotes can be found in Ref.~\cite{borish2022seeing} and some of the student quotes in Sec.~\ref{sec:contributionsToSeeingQM} also help explain the ideas presented here. 

\begin{table*}[htbp]
\caption{Codes used for both instructor and student interviews describing what seeing quantum mechanics or quantum effects means to them.}
  \centering
  \setlength{\tabcolsep}{3pt}
    \begin{tabular}{>{\hangindent=2em}>{\raggedright}p{7cm} >{\hangindent=1em}p{10.5cm}}   
    \hline
    \hline
    \textbf{Code}& \textbf{Definition}\\
    \Xhline{2\arrayrulewidth}
    \multicolumn{2}{l}{\textit{Seeing quantum effects may include...}} \\
    \hspace{0.75em} Seeing experimental results & Seeing quantum effects involves seeing experimental results or statistics. \\
    \hspace{0.75em} Clear results that require little interpretation & Seeing quantum effects involves experimental results that are very clear and involve minimal interpretation.\\
    \hspace{0.75em} Seeing and understanding experimental apparatus & Seeing quantum effects involves physically seeing parts of the experimental apparatus and/or understanding how at least part of the experimental apparatus works.\\
    \hspace{0.75em} Understanding theory behind the experiment & Seeing quantum effects involves understanding the theory behind the experiment at some point during the experimental process.\\
    \hspace{0.75em} Interactions with the experiment & Seeing quantum effects involves interacting with or doing an experiment. This encompasses both views that physically interacting with the experiment is what is important and that other kinds of interactions are what are important.\\
    \hspace{0.75em} Experiments described by quantum physics & Seeing quantum effects involves an experiment that, at least in part, can only be explained by quantum mechanics. This could be a statement about something reacting in an unexpected way or not being explained by classical physics, a comparison between classical and quantum models, or a mention of being described by quantum mechanics or proof for quantum mechanics. Quantum mechanics can show up in the experiment in different ways. \\
    \hspace{0.75em} Not literally seeing quantum objects & Acknowledgement that students do not see the photons in the single-photon experiments, so literally seeing quantum objects is not necessary for seeing quantum effects (or achieving other learning gains from working with quantum experiments).\\
   \multicolumn{2}{l}{\textit{Seeing quantum effects can help students...}} \\
    \hspace{0.75em} Believe quantum mechanics describes the physical world & Seeing quantum effects helps students realize, believe, or confirm that quantum mechanics describes the physical world (whether that is something that occurs in a lab or in their everyday lives) and make the connection between the mathematical theory and experiments that actually happen.\\
    \hspace{0.75em} Gain familiarity with quantum mechanics & Seeing quantum effects helps students build familiarity with quantum ideas, including making quantum mechanics seem less mysterious or more concrete or building intuition.\\
    \hspace{0.75em} Improve conceptual understanding & Seeing quantum effects helps improve students’ conceptual understanding of quantum mechanics.\\
    \hspace{0.75em} Learn about topics of technological and societal importance & Seeing quantum effects helps students learn about topics appearing in technological applications or pop culture. This can provide excitement or motivation for the students, help the students in their future careers, or give them the authority to be able to educate others about these topics.\\
    \hspace{0.75em} Think about philosophy of quantum mechanics & Seeing quantum effects helps students think about ideas surrounding interpretations and philosophy of quantum mechanics.\\
    \hspace{0.75em} Generate excitement and motivation & Seeing quantum effects helps students get excited and/or motivated to learn more (either in their quantum classes or to pursue physics as a career).\\
    \hspace{0.75em} Make learning quantum seem more attainable & Seeing quantum effects makes the field accessible to a variety of students. This could be because some students do not like theory/math, think of quantum as complicated or intimidating, or realize after working with a quantum experiment that it is something they can do.\\
    \hline
    \hline
  \end{tabular}
  \label{tab:seeingQMcodeDefinitions}
\end{table*}

We divided the codes related to seeing quantum effects into two main categories: one about the aspect of the experience of working with the experiment that led to students or instructors feeling like they observed quantum effects (labeled as ``seeing quantum effects may include...'' in Sec.~\ref{sec:whatIsSeeingQM}) and the other about potential benefits for students from working with the experiment (labeled as ``seeing quantum effects can help students...''). Although we made this distinction, instructors and students did not always separate these ideas. Both sets of codes came up when asked what seeing quantum effects means, as well as why it is important.

\subsection{Components that contribute to seeing quantum effects}

One component of seeing quantum effects for some students and instructors was observing some form of the results (the code \textit{Seeing experimental results}). This could be the raw data (e.g., counts from a detector recorded as either numbers or a graph, or a build up of statistics) or the data after it had been analyzed. When asked what seeing quantum effects in experiments meant to him, Casey said, ``And so I guess in a literal sense, it would just be seeing a number on a screen that shows that classical physics can't work.'' Frankie, when asked what specific parts of the experiment caused her to observe quantum effects, discussed both the ``raw sort of numbers coming out of [the detectors]'' as well as ``the values com[ing] out'' of the Mathematica and Python programs (see the first quote in Table~\ref{tab:whatContributesToSeeing}).

Some students and instructors took it a step further and claimed it was not just seeing the experimental results, but seeing results that were particularly clear and did not require additional interpretation. The code \textit{Clear results that require little interpretation} was often assigned when students and instructors compared their experiences of seeing quantum effects in different experiments within the set of single-photon experiments. Because of this, some of the student quotes were double coded with the codes in Table~\ref{tab:whatContributesToSeeing}. In particular, Sec.~\ref{sec:contributionsToSeeingQM} contains a discussion of the different ways students perceived the experimental results as clear. Instructors also discussed this idea, often with the example of how the Bell's inequality experiment involves ``many steps'' and students 
\begin{quote}
    \textit{``have to be willing to accept many things that [they’re] maybe not quite as comfortable with. Whereas if [they’ve] only got to make one leap, okay, [they] can get that, but if [they] need to make three leaps, it gets that much harder to sort of see.''} 
\end{quote}
This experiment was sometimes compared with the existence of a photon experiment, which one instructor described as ``that's really clear that they see the result and it's really clear what it means.'' 

For some students and instructors, seeing not just the results but also the experimental apparatus was important to feeling like they saw quantum effects (the code \textit{Seeing and understanding experimental apparatus}). When asked to compare the importance of seeing quantum effects with seeing other areas of physics, Morgan explained: ``quantum effects are very hard and difficult to visualize. And so seeing the setup and seeing what needs to happen is more important.'' One instructor gave the example of the physical layout being particularly important to understand the concept of superposition, because a superposition of two positions is clearer to students than a superposition of polarization states. Another instructor talked about how the students were ``closer to the quantum aspect of what was happening, because it was spread out on the table.'' 

For others, not just seeing, but also understanding how the apparatus worked was important. However, students differed in the amount of importance they attributed to it. When explicitly asked how much of the experimental apparatus they need to understand to see quantum effects, the students' responses ranged from none (the equipment was not the important part) to pointing out which parts they thought were important. For example, Alex said:
\begin{quote}
    \textit{``...the parts of equipment that are really integral to the operation of the system, you should probably know how they work and how they're affecting the system. Some of those more peripheral elements, I don't think are quite as important.''}
\end{quote}
She gave the example of the beamsplitter and polarizing films as being integral to the single-photon experiments in contrast with the power supply for the laser, which was not the focus of the lab.

In addition to understanding the apparatus, understanding the theory behind the experimental results was also part of seeing quantum effects for many students and instructors (the code \textit{Understanding theory behind the experiment}). For example, when asked if she had experimentally observed quantum effects in her course, Alex discussed a lab she had recently performed and explained how it was ``a really good example of like yes I've seen this and I understand how this is operating on a quantum level.'' Some students and instructors instead focused on having a solid understanding of classical physics. Indigo said:
\begin{quote}
    \textit{``But I think most of all, what I need to know is not necessarily the quantum concepts, but the classical concepts. Because if I have the classical concepts, then I know what to expect. And then when those expectations aren't what I see, then I know that it's something else, and obviously it'll be quantum. So I think having a classical understanding of concepts is maybe even more important than the quantum understanding in doing this kind of experiment.''}
\end{quote}

When students were asked if they needed to fully understand the concepts of the experiment to feel like they observed quantum effects, responses ranged from ``no'' to ``a hundred percent'' with many responses in between. Frankie, the student who thought it was a hundred percent necessary, further explained: ``Because otherwise it's just me looking at numbers and saying yeah they line up with some math I did that didn't make sense to me.'' Some students talked about how they think fully understanding the concepts is not necessary to observe quantum effects, but having a solid background in the theoretical side of the experiment makes it more valuable. 

The code \textit{Interactions with the experiment} was assigned to students and instructors who thought that different kinds of interactions with the experiment were necessary to observe quantum effects. Students rarely mentioned interaction being important for seeing quantum effects more generally, but it came up many times when students discussed seeing quantum effects in the context of the single-photon experiments (see Sec.~\ref{sec:contributionsToSeeingQM}). The students discussed interaction in the form of adjusting polarization optics. Some instructors defined interactions more broadly, using phrases such as ``doing an experiment'' or ``direct experimental interaction with quantum effects.'' 

The instructors were additionally asked if students could see quantum effects while watching a video of the experiments or a demonstration instead of interacting with the experiments themselves. Most of the instructors discussed how directly working with the experiment was necessary. One explained: ``But the farther you are from the experiment, the less seriously you take it as being an actual experiment versus a dog and pony show.'' They then discussed how watching a video is ``special effects'' and watching a demonstration is watching ``a magician'' before talking about how ``it's not that the students deliberately don't believe it, it's just that it's separated from the experience of I set this up myself, and I verified these are the paths that the light is taking...'' 
Another instructor explained how ``it's much better if they do the experiment'' because ``they have to spend a lot more time thinking about it... And thinking about all the details, so that they understand their results.'' 

A few other instructors were less convinced that physical interaction was completely necessary; they were open to the possibility that properly designed simulations, remote labs, or demonstrations where the students could direct the instructor might be able to improve students' understanding. For them, students being able to make decisions or change parts of the experiment was key, as evidenced when one instructor said 
``I don't think the physical interaction is as necessary as just having a large enough kind of parameter space to be able to change in the experiment.''

The fact that the experiments themselves needed to be quantum is seemingly trivial, but is also a part of many students' and instructors' ideas about seeing quantum effects. There are a variety of ways experiments could be considered quantum, and not all of them may lead to students feeling like they observe quantum effects, as is discussed in Sec.~\ref{sec:quantum-waysManifests}. The code \textit{Experiments described by quantum physics} was assigned to students and instructors who talked about how the experiments needed to be quantum. 

In addition to the ideas in discussed in Sec.~\ref{sec:quantum-waysManifests}, one of the main ways students and instructors brought this up was by explaining that they see quantum effects when an experiment can be explained by a quantum, not a classical, model. One instructor said:
\begin{quote} 
    \textit{``I try to emphasize not so much seeing quantum mechanics... I really try to emphasize what's the difference between classical physics and quantum physics...  a big thing is trying to draw that line between what can we explain classically using Newtonian physics or whatever, and then... when is it absolutely necessary to use quantum mechanics.''}
\end{quote}
Casey discussed the experiments in a similar way. When asked what it means to him to see quantum effects in experiments, he said:
\begin{quote}
    \textit{``... it was mainly showing that there were certain quantities that had I guess different predictions. If you use classical versus non classical models... and so the entire experiment was usually... about showing that those quantities would lie in a non classical regime. And so I guess in a literal sense, it would just be seeing a number on a screen that shows that classical physics can't work... you formulate some test to show that... quantum mechanics provides a more accurate prediction, and then you explicitly show that.''}
\end{quote}
Some students, however, were less explicit and instead discussed experiments in relation to what they expected based on their classical experience with the world. For example, Nicky talked about ``when stuff doesn't accord with what we think would happen, or what like classical mechanics tells us would happen,'' and Hayden talked about ``things not reacting the way that they should.''

Lastly, some students and instructors brought up the fact that even though they were talking about seeing quantum effects, it is not possible to actually see the photons themselves (the code \textit{Not literally seeing quantum objects}). This was always discussed in the context of the single-photon experiments, so student ideas about this code are briefly discussed in Sec.~\ref{sec:contributionsToSeeingQM-differentCombinations}. One instructor, for whom this led to skepticism about the idea of seeing quantum effects, said: ``yeah I mean I am also kind of skeptical about this idea of seeing quantum mechanics because you don't see the beams, right. They're single photons, you don't see them.'' Another instructor discussed how they wished students could see more: ``Yeah, it would be... a much nicer lab in my mind if you could see, if the down converted beams were bright enough to see.'' Another instructor instead talked about ways students could understand they were working with photons even if they could not physically see them. They described the photons as ``stuff you can't see with your eyes... And yet, you can move knobs and get signals from it.'' 

\subsection{Learning goals achievable by seeing quantum effects}

One of the frequent codes for both students and instructors was \textit{Believe quantum mechanics describes the physical world}. This code encompasses primarily two intertwined ideas. The first is that quantum mechanics is not just math or theory; it is what happens in the physical world. When asked what it meant to her to see quantum effects in experiments, Frankie discussed this idea:
\begin{quote}
    \textit{``Quantum is like pretty abstract, for the most part, when you're learning about it. It's just like this thing that happens. And classical is much easier to get behind because it's something you can see. You see it every day. But this is like a really good opportunity to actually see quantum physics in play. And so it just sort of like grounds it a little more in reality, instead of just being this vague like, I don't know, sort of thought experiment, for the most part. So yeah I like just being able to see the physics happening. Or just having some like actual visual proof.''}
\end{quote}
Others talked about this as a math--physics connection. Although many students thought quantum mechanics was taught in a mathematical way, Kai brought up how at his institution, quantum is framed as being experimental, which makes it particularly important to see in an experimental setting:
\begin{quote}
    \textit{``And the way [quantum is] taught is that it should be confusing in that it doesn't have answers, and that it's very experimental, and that everything we're going off of is experimental evidence. But it's hard to connect with that and `understand' that if you don't get to see how it's `just experimental evidence.' ''}
\end{quote}

The second main idea encompassed by the code \textit{Believe quantum mechanics describes the physical world} is that seeing something helps students believe it more than just being told about it. For example, when asked if it is important to see quantum effects, Greer said:
\begin{quote}
    \textit{``I think it was... very important to see that the stuff that we're talking about that sounds unrealistic as you're first learning it, like things can exist in two states, but being able to see that the effects in the lab really just say no, this is really what's going on. And just kind of... adds confidence to what you're learning. And makes it a little bit easier to learn, because you're like okay... I've seen it work now. I can more easily accept that this is the way it works.''}
\end{quote}
As discussed in Sec.~\ref{sec:notJarring}, the students already believed in quantum mechanics, but working with experiments still helped confirm their beliefs. Indigo explained: ``When I get to college, I'm already not in the level that I need to see things in order to believe them, but it's nice to get a confirmation...''

Another common code for both students and instructors was \textit{Gain familiarity with quantum mechanics}. Just as with the previous code, this one encompasses several related ideas and could be considered a bridge between the codes about believing that quantum mechanics describes the physical world and conceptual understanding. One of the ideas contained within this code is that seeing quantum effects can help students build intuition. For example, when asked to compare seeing quantum effects with seeing experimental effects from other areas of physics, Dana said:
\begin{quote}
    \textit{``In almost all other areas of physics, we do see it in our everyday life as well. We have an intuition for like kinematics and motion and somewhat of an intuition for light and how it'll react. But we don't see quantum effects at all. So, it's like I feel like that's the best thing to do in a lab is look at things that you don't normally get to see.''} 
\end{quote}
Although the concept of intuition was brought up often by both students and instructors, intuition has different meanings for different students and can also be related to the math--physics connection \cite{corsiglia2023intuition}. 

Another part of the code \textit{Gain familiarity with quantum mechanics} is that seeing quantum effects can make the field seem more concrete since it is often perceived as abstract, intangible, and inaccessible. When asked if it was important to see quantum effects in experiments, Logan said:
\begin{quote}
    \textit{``I think it makes... a field that oftentimes seems intangible, because it's... generally such a small scale that you need very high end equipment to see it. That it's inaccessible to a lot of people, so it makes it feel more real and less like an esoteric concept that's in a class.''}
\end{quote}
Others instead focused on how weird or mysterious quantum mechanics is. When asked what seeing quantum effects means to him, Briar talked about quantum as being ``super weird... because it sort of goes against my intuition and everything that I think about the way the world works.'' Seeing quantum effects in an experiment can help students become more familiar with the abstract and seemingly weird concepts.

Another reason instructors and students want students to see quantum effects is to help them learn concepts (the code \textit{Improve conceptual understanding}). Concepts is a broad phrase that can include many different ideas, so for our coding scheme in Sec.~\ref{sec:learningGoals}, we created sub-codes for improving understanding of quantum concepts (e.g., particle-wave duality or entanglement), the apparatus, and uncertainty and statistics. The third category was included since quantum mechanics is not deterministic, so students need a solid basis in probability, statistics, and the way uncertainty is different in quantum mechanics than classical mechanics \cite{bao2002understanding, stein2019student, white2020student, stump2020student}. This code, however, focuses primarily on the first of these three categories, since students often can learn how parts of the apparatus work in non-quantum experiments and discussions of statistics unique to quantum mechanics were less frequent. 

Instructors discussed learning about quantum concepts in the context of seeing quantum effects in more specific ways than the students did. Students talked about conceptual learning with only vague words, such as ``better understanding.'' Instructors went into more details, including how there are some concepts that are difficult to understand entirely theoretically. For example, one instructor 
said: 
\begin{quote}
    \textit{``...there are some things like entanglement which I feel like they're really difficult to understand just on the basis of math, manipulating mathematical formula. And so I think it's helpful to see a lab where you're seeing a result which you can only understand on the basis of entanglement.''}
\end{quote} 
Other instructors talked about about how seeing quantum mechanics could help students get rid of misconceptions: ``And so seeing quantum mechanics could be a way to to defeat some of the wrong things that one might think about what an entangled state is.''

Some of this learning can have broader implications outside of the physics classroom as well, as is evidenced by the code \textit{Learn about topics of technological and societal importance}. Some students discussed the technological side of this. They knew that quantum experiments could be related to quantum technologies, even if they did not fully understand how quantum technologies worked. For example, when describing what seeing quantum effects in experiments meant to him, Greer said:
\begin{quote}
    \textit{``Just seeing [effects] in the experiment shows me that they could have and will have potential beyond just the niche physics experiments in the lab. And that there definitely are ways to, or there's probably ways to utilize it in the future in a more macro, daily scale. And whether that's through quantum computers or something else, I don't know yet.''}
\end{quote}
Greer is referring to an example of a quantum 2.0 technology, the more recent quantum technologies, such as quantum computing, that utilize the manipulation of quantum entanglement \cite{dowling2003quantum}. Some instructors discussed seeing quantum mechanics as helping students learn about quantum 2.0 technologies, whereas others focused on students seeing quantum experiments from the previous generation of quantum technologies. This second category, which includes semiconductor physics, may be more relevant for their engineering students to see.

The second component of the code \textit{Learn about topics of technological and societal importance} is related to the way quantum mechanics shows up in society at large. This idea was not mentioned by the students. Instructors talked about how students may ``appreciate being able to say something about [quantum ideas].'' They could even use this knowledge to teach others. One instructor 
talked about how students can ``be stewards'' and answer questions people in the public have about quantum physics. They went on to explain how after working with quantum experiments, students will 
\begin{quote}
    \textit{``have authority to say I've gone through that from theory and shown it in experiment and I've seen, it's not just something that somebody is saying and writing down on a blackboard or whiteboard, it's something that I've seen in the lab.''}
\end{quote} 
The knowledge students gain from quantum experiments can also be used to combat misinformation. When talking about ways entanglement is often misused, one instructor discussed performing Bell's inequality, which is a proof of entanglement:
\begin{quote}
    \textit{``We need to go out of our way, we go into the lab, we use lasers, we use spontaneous parametric down-conversion. Like you can't, there's no way that I know of to entangle a vaccine with the Google credit score... if you realize that you have to spend hours and hours for three weeks to achieve this entanglement... they can use our words, like we don't have to be possessive of our words, but when they say quantum entanglement it's not what we mean.''}
\end{quote}

Although less commonly discussed, one theme that appeared in some instructor interviews is the code \textit{Think about philosophy of quantum mechanics}. This often related to the various interpretations of quantum mechanics, and how seeing an experiment can help students think about them. When asked what seeing quantum mechanics means to them, one instructor 
said: 
\begin{quote}
    \textit{``For something like the entangled photons, I think it has a lot to do with interpretations of quantum mechanics, because if we're visualizing, we're sort of imposing on the photons some kind of state prior to measurement.''}
\end{quote} 
Doing an experiment can lead to thinking about the state of the photons throughout the entire experimental sequence, which can differ based on the favored interpretation. Other instructors also mentioned how there are ``different ways to conceptualize [what's happening]'' or how an important idea in these experiments is ``whether or not you view photons as real physical objects or as artifacts of measurement.'' None of the students brought up philosophy with respect to seeing quantum effects.

Seeing quantum effects can have an affective element as well (the code \textit{Generate excitement and motivation}). This could come about through student excitement about the experiments themselves or the experiments helping motivate students to complete their coursework and pursue physics in the future. When asked what it meant to students to see quantum effects in experiments, one of the first things some of the students said was ``it's really cool.'' There is a large range of reasons these experiments may be exciting for students, including the concepts \cite{borish2023implementation}. In relation to this idea of seeing quantum mechanics, some instructors wanted their students to have the ``feeling of wow I just saw magic'' or ``view it as this hidden knowledge.'' Excitement caused by these experiments could also help students be motivated to spend time understanding their coursework or study more physics in the future. For example, Jaime discussed how it's ``very important to see quantum effects because they are what motivate new physics... So, I think those are important to really motivate people to study more physics, more than the basic level.''

In addition to motivation, some instructors and students think these experiments can improve students self-confidence in their ability to understand quantum mechanics (the code \textit{Make learning quantum seem more attainable}). Students talk about how ``you hear from sources that like quantum physics is so hard'' or how they ``have classmates that basically are like I get it, but it's too hard, it intimidates me, I don't want to do it.'' Instructors discuss how seeing quantum mechanics can be an entry point for some students to feel that they are capable of understanding it. For students who are intimidated by the math, experiments can be a way for them to still experience quantum mechanics without all of the associated math. One instructor said: ``And I think this system... opened up quantum mechanics for folks who were more interested in the applications and who liked more experimental physics.'' Additionally, having experience working with a quantum experiment that is similar to some kinds of current research can make future research opportunities ``more accessible'' to the students because ``[they]'ve talked about quantum in modern physics or [they]'ve taken the quantum course and this is something [they] can do.''

\bibliography{seeingQM}

\clearpage


\section*{Supplementary material}

# Seeing quantum effects in experiments
## Supplemental Material

In this supplement, we include the parts of the interview protocols relevant for the data presented in this work for both instructors and students.

## A. Instructor interview protocol

This section contains the questions related to seeing quantum mechanics that we asked instructors. These were semi-structured interviews, so the exact wording of the questions may have varied for each instructor and some questions may have been combined if there was not sufficient time. These questions were preceded by questions about the course(s) the instructor teaches and followed by questions about other learning goals the instructors had for using the single-photon experiments.

1. Before moving on to discuss the single-photon experiments, I want to focus more generally on the idea of seeing quantum mechanics. What does the term seeing quantum mechanics mean to you?

2. Now moving on to thinking about your course, what does the term seeing quantum mechanics mean to you for the single-photon experiments?

3. Do you use the phrase seeing quantum mechanics with your students, and if so, how do you discuss it with them?

4. In the survey, you responded that the goal of "seeing quantum mechanics in real life" was [their survey response]. Why do you think it's important for students to see quantum mechanics in a physics class?

5. Which quantum concepts (e.g., wave-particle duality, entanglement, superpositions) do you think are most important for students to see in lab? Why?

6. Do you think it's more, less, or equally important for students to see quantum mechanics compared with seeing other areas of physics?

7. What do the students in your course think seeing quantum mechanics means?

8. Do you think students in your course see quantum mechanics while working with the single-photon experiments?

9. What specific parts of these experiments cause students to see quantum mechanics?

10. Do you think students see quantum mechanics more in some of the experiments in your course than others? If so, what it is about those experiments that causes this?

11. Do you think students see quantum mechanics when they watch a video of a single-photon experiment?

12. Do you think students see quantum mechanics when they see the physical set-up (not a video) and watch someone else perform a demonstration of a single-photon experiment?

13. Do you think students need to physically interact with or manipulate in some way the single-photon experiments to see quantum mechanics?

14. Do you think the students themselves feel like they see quantum mechanics while working with the experiments?

## B. Student interview protocol

This section contains the relevant parts of the student interview protocol. These were again semi-structured interviews, so the exact wording of the questions may have varied slightly by student based on their prior responses. These questions were preceded by questions about the student's prior experiences with quantum mechanics and research and were followed by questions about other learning goals and a few demographic questions.

1. What does it mean to you to see quantum effects in experiments?

2. Do you think it's important to see quantum effects in experiments?

3. Do you think it's more, less, or equally important to experimentally see quantum effects in a lab compared with seeing other areas of physics?

4. I will ask about your course from this semester later on, but do you feel like you've observed quantum effects experimentally in any of your prior courses?

5. Do you feel like you experimentally observed quantum effects during your course? *If so:* Which effects did you observe?

6. Do you think you experimentally observed quantum effects more in some of the experiments in your course than others? Why?

7. What specific parts of the experiment caused you to observe these quantum effects?

8. Do you need to fully understand the concepts of the experiment to feel like you observe quantum effects?

9. Do you need to fully understand the different pieces of equipment to feel like you observe quantum effects?

10. Did these experiments feel like a black box to you (by that, I mean that how everything worked was mysterious and you turned on the laser and then obtained results without knowing what happened in the middle) or were you able to understand what each of the different parts of the experiment did?

11. Did working with these experiments help you better understand quantum concepts that you've learned about in classes or through readings or pre-labs assignments?

12. Do you see a connection between the math you've studied in your theory classes and/or read about in pre-labs and the experimental results you found with these experiments? If so, what about these experiments helped you make that connection?

13. Did working with these experiments help you build intuition about quantum effects?

14. Did working with these experiments change your beliefs about the validity of any quantum effects or quantum mechanics as an entire field? If so, what caused that belief to change?

15. Did you ever have an "aha" moment while working with these experiments? If so, can you describe that moment?

16. Did you ever have an "aha" moment while learning about the concepts covered in this experiment but in a theory class?

17. Do you understand what is quantum about these experiments and how it differs from classical ideas?

18. Have your views about whether or not quantum mechanics is weird or mysterious changed after working with these experiments? If so, how?

19. Before taking this course, could you explain how to experimentally create quantum entanglement or quantum superpositions?

20. After taking this course, can you now explain how to experimentally create quantum entanglement or quantum superpositions?

21. Did these experiments make you think about the purpose of why we do experiments in physics?

22. Did you find working with these experiments to be exciting or motivating? If so, what about these experiments did you find particularly exciting/motivating?

23. Did you find working with these experiments to be frustrating? If so, what in particular was frustrating? Did you learn anything from those frustrating moments?

24. Do you see a connection between these experiments and quantum technologies such as quantum computing or quantum cryptography?

25. Do you think your experience working with these experiments will be useful for you in the future (either in other physics classes or a future job or any other aspect of life)?

26. What pieces of equipment did you learn how to use while working with these experiments?

27. Did you learn something about uncertainty and statistics while working with these experiments? If so, is it similar or different to what you've learned while working with other experiments in physics lab courses?

28. Is there anything else you would like to share with me about your experiences working with these experiments?



\end{document}